\documentclass[twocolumn]{article}

\usepackage{amsmath}
\usepackage{amssymb}
\usepackage{graphicx}

\newcommand{\helptext}[1]{}
\newcommand{\marginhelp}[1]{}

\newcommand{\be}{\begin{equation}}
\newcommand{\ee}{\end{equation}}

\newcommand{\de}{\stackrel{\mbox{\tiny def}}{=}}

\newcommand{\plabel}[1]{_{\{#1\}}}

\newcommand{\mygraphics}[1]{\includegraphics[width=2.3in]{#1}}

\begin{document}

\title{Cash Flow Entropy}
\author{Ulrich Kirchner \\[0.25cm]
\small PO Box 72277, Parkview, 2122, South Africa \\ \small ulrich.kirchner@gmail.com\\[1cm]
Simon Moolman \\[0.25cm]
\small Department of Physics\\ \small University of the Witwatersrand\\ \small Johannesburg\\[1cm]
}

\maketitle

\begin{abstract}
In this paper we aim to find a measure for the diversity of cash flows between agents
in an economy.
We argue that cash flows can be linked to probabilities of finding a currency
unit in a given cash flow. We then use the information entropy as a natural measure of
diversity.
This leads to a hirarchical inequality measure, which includes the well-known Theil index as a special case,
and a constraint on agent entropy differentials, i.e., the difference between cash inflow and outflow entropies.
The last result is particularly intriguing as it formalises the fact that an economy must contain both,
cash flow aggregating and cash flow diversifying agents.
\end{abstract}

\section{Introduction}

Most people, when inspecting their bank account statement, could make the following observation: while
inflows tend to come in a few large payments (salary etc), outflows are much more dispersed (daily groceries, rent,
monthly bills, etc).
By considering the sources, destinations and relative sizes of the payments, it is possible to write down the probability of where a dollar of an agent's income came from and where it is going.
With this we can describe the difference between the large inflows and dispersed outflows in terms of the entropy of these probabilities.

Let us assume an economy in which no money is created or destroyed and in which, over a period, cash flows only occur between a set of discrete agents.
Such an economy can clearly not exist if every agent's cash flow situation is as above --- few large inflows and many small outflows.
There must be agents who collect small cash flows and pay out large sums.
Examples which come to mind are generally companies, for example selling a mass produced product (small cash flows) and buying supplies
and services in bulk (large payments).

This hints at the idea that the entropies of the inflow and outflow probabilities of all the agents in the economy must be related in some way.

This paper will be structured as follows: In Section \ref{sec2} we formalize the cash flow-probability relationship and
introduce information entropy to measure diversity. In Section \ref{sec3} we examine the results
for the simpler case of a steady-state economy. Section \ref{sec4} contains examples for the two and three agent cases.
This is followed by the Conclusion in Section  \ref{sec_conc}. In the appendix we derive the information entropy subdivision property for the case of 2/3 possible
outcomes.

\section{Cash flows and probabilities}
\label{sec2}

Let us assume that in an economy cash flows occur over a period of time between $N$ agents.  Let $c^j_k$ represent the cash flow from agent $j$ to agent $k$.
This can be conviniently represented in a cash flow matrix, where each row corresponds to a source agent (upper index)  and each column to a destination agent (lower index)
\be
\left[
{\begin{array}{ccc}
c^1_1 & \cdots & c^1_N \\
\vdots & & \vdots \\
c^N_1 & \cdots & c^N_N
\end{array}}
\right]
\ee
With cash flow $c^j_k$ we can then associate a probability
\be
p^j_k \de \frac{c^j_k}{\sum_{k,j} {c^j_k}}.
\ee 
which
represents the likelihood that a random currency unit in the economy would have been part of cash flow $c^j_k$ (which assumes identifiable currency units).
Note that ``saving'' would correspond to a cashflow $c^j_j$ from an agent $j$ to himself. 

To measure the diversity of all cash flows in the economy we will use the information entropy
\be
H \de - \sum_{j,k} p^j_k \log_2(p^j_k),
\ee
which is measured in bits (due to the choice of $2$ as the base of the logarithm).
To simplify the notation let us extend this definition to ``un-normalized'' probabilities by defining a new function
\be
K(x_1, \ldots, x_n) \de H\left(\frac{x_1}{\sum_i x_i}, \ldots,\frac{x_n}{\sum_i x_i}\right).
\ee

An intriguing property of  the information entropy is how it respects the grouping of possible outcomes.
Lets say we group the $N$ possible outcomes into $n<N$ groups. The overall entropy can then be written as 
\be
H= H_{\mathrm g} + \sum_j p_j H_j,
\label{eq-50}
\ee
where $p_j$ is the probability of the outcome being in group $j$, $H_j$ is the ``internal entropy'' of group $j$ (entropy of the probabilities conditional on the fact that the
outcome was in group $j$), and
\be
H_{\mathrm g}\de - \sum_j p_j \log_2 p_j
\ee
is the information entropy for the aggregate groups.

By applying above identity recursively to the probabilities corresponding to all possible cash flows one obtains a hirarchical set of inequality/diversity measures.

As a first step let us put all savings $c^i_i$ in one group, and all remaining cash flows in another group.
Using above identity we find
\be
H = H\plabel{sc} + p\plabel{s} H\plabel{s} +p\plabel{c} H\plabel{c}, 
\ee
where $p\plabel{s}\de \sum_i p^i_i$ and $p\plabel{c}=1-p\plabel{s}$ are the aggregate probability of a currency unit being in a ``saving cash flow'' and inter-agent cash flow respectively,
\be
H\plabel{sc} \de -p\plabel{s} \log_2 (p\plabel{s}) - p\plabel{c} \log_2 (p\plabel{s}) 
\ee
is a measure of diverisification between savings and spending,
\be
H\plabel{s} \de K(c^1_1, \ldots, c^N_N)
%H\plabel{s} \de -\sum_j \frac{c^j_j}{\sum_k c^k_k} \log_2\left(\frac{c^j_j}{\sum_k c^k_k}\right) 
\ee
is a measure of the saving diversity in the economy (it is highest when all agents have the same saving), and
\be
H\plabel{c} \de K(c^1_2, \ldots, c^n_{n-1})
%H\plabel{c} \de - \sum_{j\ne k} \frac{c^j_k}{\sum_{m\ne n} c^m_n } \log_2\left(\frac{c^j_k}{\sum_{m\ne n} c^m_n } \right) 
\ee
is a measure of diversity for all inter-agent cash flows.

%We can express the overall entropy in terms of the entropy between groups (the agents) and an internal entropy
%\be
%H= H' + \sum_j p^j H^j,
%\ee
%where $H'$ is the entropy of the aggregate agent cash out-flow probabilities $p^j$ and $H^j$ is the entropy of the $p^{(j)}_k$ for fixed $j$. 

Cash flows can now be grouped by their origin or destination. We define the aggregated cash flows {\em from} agent $j$
and {\em to} agent $k$ {\em excluding savings} as
\be
c^j \de \sum_{k, k\ne j} c^j_k
\qquad
c_k \de \sum_{j,j\ne k} c^j_k.
\ee

The probabilities of finding a currency unit (which is not a saving) in the cash-out/in flows of an agent are
\be
p^j= \frac{c^j}{\sum_i c^i}
\qquad
p_k= \frac{c_k}{\sum_l c_l}
\ee

We can now define probabilities similar to above, but within one of the cash flow groups {\em to} or {\em from} an agent for {$j \ne k$}
\be
p^{(j)}_k = \frac{c^j_k}{c^j}(1-\delta^j_k)
\quad
p^j_{(k)} = \frac{c^j_k}{c_k}(1-\delta^j_k).
\ee
This is the probability to find an arbitrary currency unit leaving/going to agent $j$ in a cash flow to/from agent $k$.
Evidently
\be
\sum_k p^{(j)}_k =1
\qquad
\sum_j p^j_{(k)} =1
\ee

By grouping the cash flow by the originating agent (upper index) we find
\be
H\plabel{c}= H' + \sum_j p^j H^j,
\ee
where
\be
H' \de - \sum_j p^j \log_2 (p^j)
\ee
is the entropy of the aggregate agent cash out-flow probabilities $p^j$ and
\be
H^j \de - \sum_k p^{(j)}_k \log_2 (p^{(j)}_k)
\ee
is the entropy of each agents cash outflows, i.e., the entropy of $p^{(j)}_k$ for fixed $j$.
It is maximized if all cash flows from the agent are equal, and minimized (in which case it is zero) if there is only a single cash flow from agent $j$.
Hence this is a measure of spending diversity for agent $j$.

Similarly this can be done for cash in-flows
\be
H\plabel{c}= H'' + \sum_k p_k H_k,
\ee
where $H''$ is the entropy of the aggregate agent cash in-flow probabilities $p_k$ and $H_k$ is the entropy of the $p^{j}_{(k)}$ for fixed $k$. Here similarly to above
\be
H'' \de - \sum_k p_k \log_2 (p_k)
\ee
\be
H_k \de - \sum_j p^{j}_{(k)} \log_2 (p^{j}_{(k)}).
\ee

Taking the sum and difference of the two expressions for $H\plabel{c}$ we find
\be
H\plabel{c} = \frac{H' + H''}{2} + \frac{1}{2}\sum_j(p_jH_j + p^j H^j)
\label{eq-210}
\ee
\be
0=\frac{H'' - H'}{2} + \sum_j(p_jH_j - p^j H^j)
\label{eq-220}
\ee
The last equation is particularly interesting as it relates the difference of overall income and spending inequality to individual agents cash flow diversification.

\section{The Steady State Economy}
\label{sec3}

Cash flows to and from agents over the period are related by
\be
[ c^1 \cdots c^N] \left[
{\begin{array}{ccc}
p^{(1)}_1 & \cdots & p^{(1)}_N \\
\vdots & & \vdots \\
p^{(N)}_1 & \cdots & p^{(N)}_N
\end{array}}
\right]
=[ c_1 \cdots c_N]
\ee

For a stationary wealth distribution we need $c_j=c^j$ and hence
\be
p_j=p^j.
\ee
Assuming stationarity and dividing above equation by the sum of all cash flows gives
\be
[ p^1 \cdots p^N] \left[
{\begin{array}{ccc}
p^{(1)}_1 & \cdots & p^{(1)}_N \\
\vdots & & \vdots \\
p^{(N)}_1 & \cdots & p^{(N)}_N
\end{array}}
\right]
=[ p^1 \cdots p^N]
\ee
%The matrix of probabilities is a ``probability matrix'' and it is guaranteed to have an eigenvector with components of equal sign (which can then be choosen positive) and with eigenvalue $1$.
%If the matrix is irreducible then the stationary state is naturally approached in the long run.

In the case of stationarity we have $\bar H \de H' = H''$ and  expression \ref{eq-210}  becomes
\be
H = \bar H + \frac 1 2 \sum_j p^j(H_j + H^j),
\ee
which shows how the overall entropy splits into an inter-agent component $\bar H$ and an internal component for each agent $\frac{H_j + H^j}{2}$.
We suggest that $\bar H$ is a measure of income inequality as it measures the diversity of incomes.
However, $\bar H$ includes contributions from all agents, including non-human agents like corporations and government.
Using the same identity as above one can express $\bar H$ in terms of agent groups, for example
\be
\bar H = H^* + p^*_G H^*_G + p^*_C H^*_C + p^*_P H^*_P,
\label{eq-270}
\ee
where $H^*_P$ measures the income distribution between human individuals and can be recognized as related to the Theil index of income inequality \cite{theil}.
$H^*_G$ and $H^*_C$ are similar measures for government and corporates.
$H^*$ then measures the inequality between the sectors in the economy.
It is well know that these properties of the information entropy make it a suitable inequality measure \cite{shorrocks}.

Similarly (\ref{eq-220}) simplifies to
\be
0 = \sum_j p^j (H_j -H^j),
\label{eq-280}
\ee
which says that the probability weighted sum of the difference of agent inflow and outflow entropy vanishes.
This equation is interesting as it shows how the entropy differentials between agent in and out flows have to balance in the economy.

For the agent in the economy labelled $j$, the entropy $H_j$ represents the uncertainty of which source a dollar of $j$'s came from.
Similarly, the entropy $H^j$ is the uncertainty of where a dollar of $j$'s payments will go. The difference $(H_j -H^j)$ is, therefore, the difference of uncertainty between the inflow and outflow probabilities.
If the difference is positive, then an agent will be more uncertain about where an inflow dollar came from than where an outflow dollar is going. If the difference is negative then the opposite will be true.

Since the probabilities $p^j$ in (28) are all positive, it is not possible for all agents to have a positive entropy difference or all to have a negative entropy difference.
Some agents must be more certain about where their next dollar will come from relative to where it is going and other agents must be more certain about where their next dollar is going than where it came from.
Typically, the former could be an individual (often the whole income comes in the form of a single salary payment)
while the latter could be a store or service provider (many customers pay for goods while the payments will go to few suppliers).

\section{Examples}
\label{sec4}
\subsection{The two-agent economy}

This is the simplest possible, but somewhat restricted case. Let us define the matrix of cash flows
\be
C \de \left[
{\begin{array}{cc}
c^1_1 & c^1_2 \\ c^2_1 & c^2_2
\end{array}}
\right]
\ee
Note that there is the freedom to rescale the cash flows by a constant factor.
For stationarity we require
\be
c^1_2 =c^2_1.
\ee
With
\be
\vec u \de \left[
{\begin{array}{c}
1\\ 1
\end{array}}
\right]
\ee
we have
\be
\left[
{\begin{array}{c}
c^1\\ c^2
\end{array}}
\right]
=C \vec u
\ee

\be
\left[
{\begin{array}{c}
p^1\\ p^2
\end{array}}
\right]
=\frac{C \vec u}{ \vec u' C \vec u}
\ee
\be
\left[
{\begin{array}{cc}
p^{(i)}_1 & p^{(i)}_2
\end{array}}
\right]
=
\left[
{\begin{array}{cc}
c^i_1 & c^i_2
\end{array}}
\right]
/(c^i_1 + c^i_2)
\ee

From these all the entropies of interest ($H'$, $H_j$ and $H^j$) can be calculated.

Let us scale the cashflows such that $c^1_2=c^2_1=1$, i.e., the savings are measured relative to the inter-agent cashflows.
The cashflow matrix then takes the form
\be
C \de \left[
{\begin{array}{cc}
a & 1 \\ 1 & b
\end{array}}
\right],
\ee
where $a$ and $b$ are non-negative numbers.
We then have
\be
p\plabel{c}=\frac{2}{2+a+b}
\qquad p\plabel{s}=1-p\plabel{c}
\ee
\be
H\plabel{s}=K(a,b)
\qquad H\plabel{c}=1
\ee

\begin{figure}
\begin{center}
\mygraphics{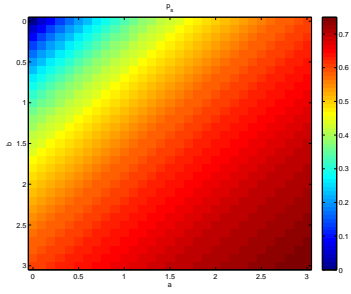}
\end{center}
\caption{
Heat-map of $p\plabel{s}$ for the 2-agent case.
}
\label{fig-2D-ps}
\end{figure}

\begin{figure}
\begin{center}
\mygraphics{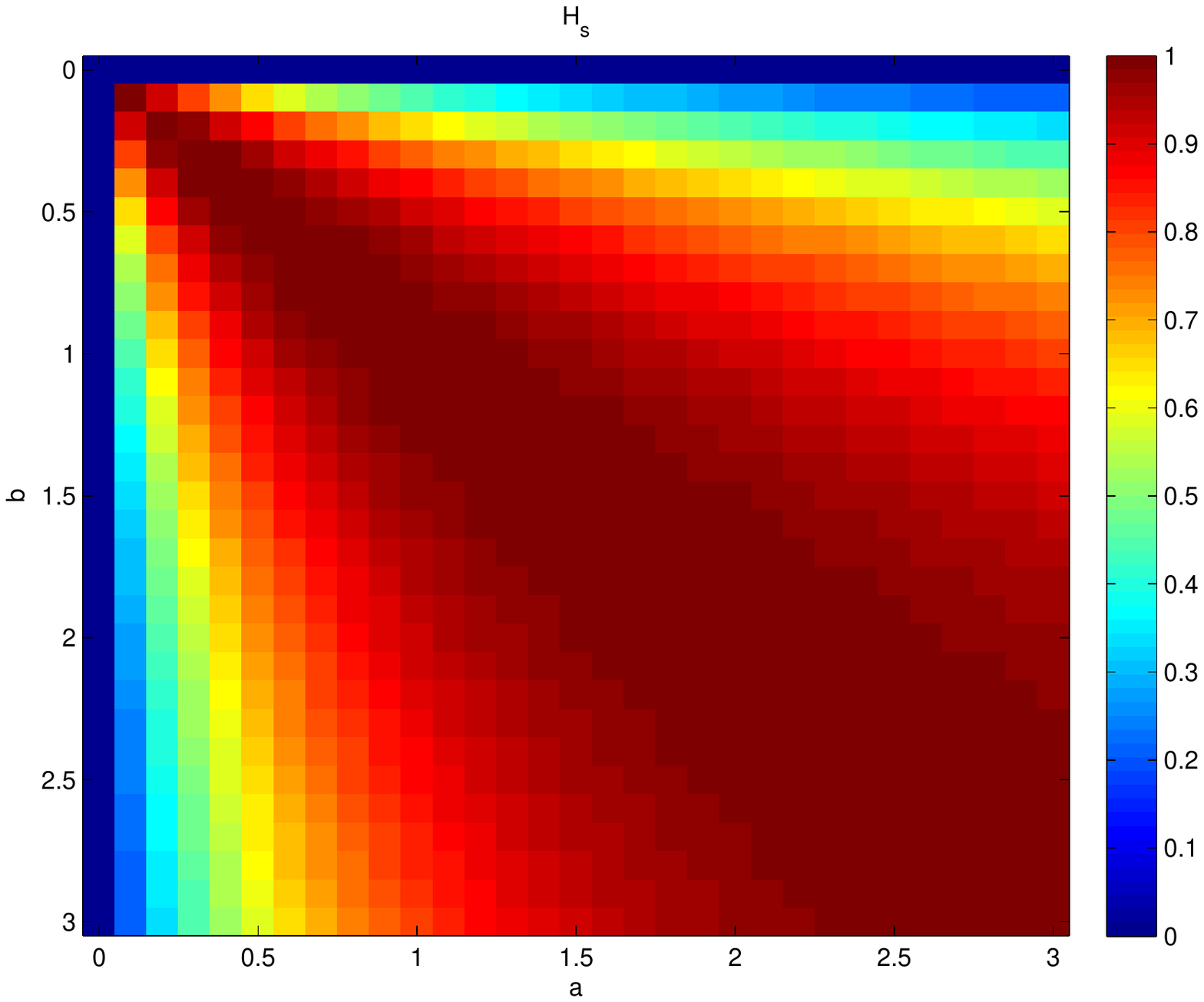}
\end{center}
\caption{
Heat-map of $H\plabel{s}$ for the 2-agent case.
}
\label{fig-2D-Hs}
\end{figure}

\begin{figure}
\begin{center}
\mygraphics{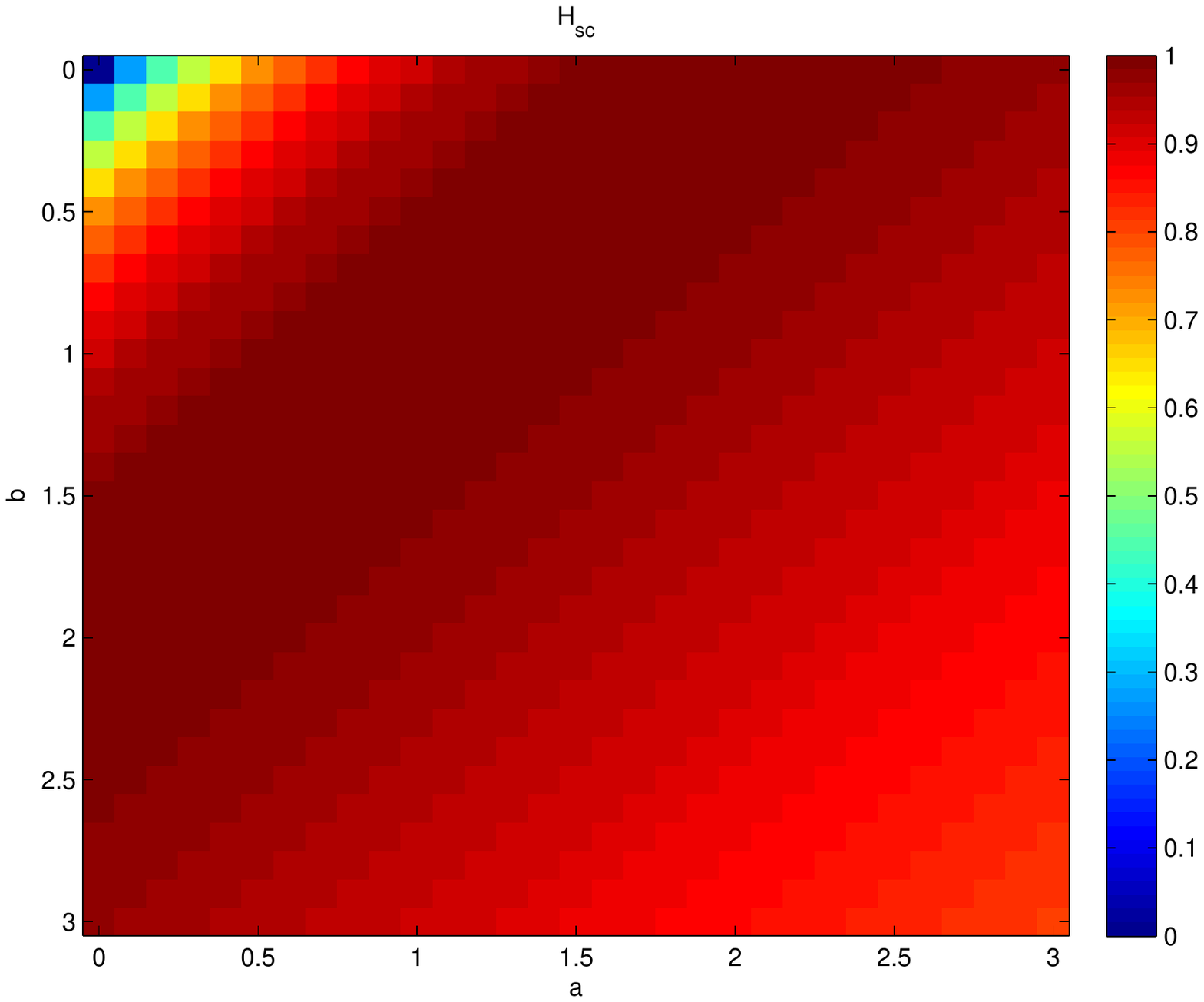}
\end{center}
\caption{
Heat-map of $H\plabel{sc}$ for the 2-agent case.
}
\label{fig-2D-Hsc}
\end{figure}

\begin{figure}
\begin{center}
\mygraphics{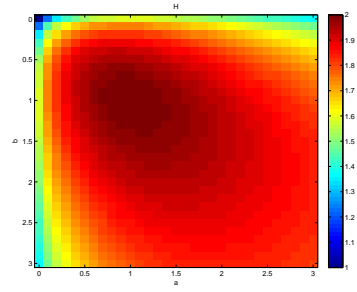}
\end{center}
\caption{
Heat-map of the overall cash-flow entropy $H$ for the 2-agent case. The previous plots are a decomposition of this entropy. We have
$H=H\plabel{sc} + p\label{s} H\plabel{s} +p\plabel{c} H\plabel{c}$.
}
\label{fig-2D-H}
\end{figure}

Figures \ref{fig-2D-ps} to \ref{fig-2D-H} illustrate probabilities and entropies for this case.

\subsection{The three-agent economy}

Let us assume stationarity and no savings. The cash-flow matrix is then
\be
\left[
{\begin{array}{ccc}
0 & c^1_2 & c^1_3 \\
c^2_1 & 0 & c^2_3 \\
c^3_1 & c^3_2 & 0 
\end{array}}
\right]
\ee
There are two constraints resulting from the stationarity requirement
\be
c^1=c_1 \qquad c^2=c_2
\ee
(this implies $c^3=c_3$) and one freedom to ``choose the cash denomination'' (allowing us to re-scale all values in the cash-flow matrix).

Let us choose the normalization such that the average agent income/spending is 1, i.e.,
\be
c^1+c^2+c^3=3.
\ee
Furthermore, let us choose the following three parameters for the cash flow matrix
\be
\frac{c^1_2}{c^1}= a \qquad \frac{c^2_3}{c^2}=b \qquad \frac{c^3_1}{c^3}=k.
\ee
Then the definition of $c^j$ implies
\be
\frac{c^1_3}{c^1}= 1-a \qquad \frac{c^2_1}{c^2}=1-b \qquad \frac{c^3_2}{c^3}=1-k.
\ee
Writing down the definitions for $c^1$, $c^2$ and the mean constraint as a linear system in the $c^j$ gives
\be
\left[
{\begin{array}{c}
c^1\\
c^2\\
3
\end{array}}
\right]
=\left[
{\begin{array}{ccc}
0 & 1-b & k \\
a & 0 & 1-k \\
1 & 1 & 1 
\end{array}}
\right]
\left[
{\begin{array}{c}
c^1\\
c^2\\
c^3
\end{array}}
\right]
\ee
and hence
\be
\left[
{\begin{array}{c}
c^1\\
c^2\\
c^3
\end{array}}
\right]
=
B^{-1} \left[
{\begin{array}{c}
0\\
0\\
3
\end{array}}
\right],
\ee
where
\be
B=
\left[
{\begin{array}{ccc}
-1 & 1-b & k \\
a & -1 & 1-k \\
1 & 1 & 1 
\end{array}}
\right]
\ee

\begin{figure}
\begin{center}
\mygraphics{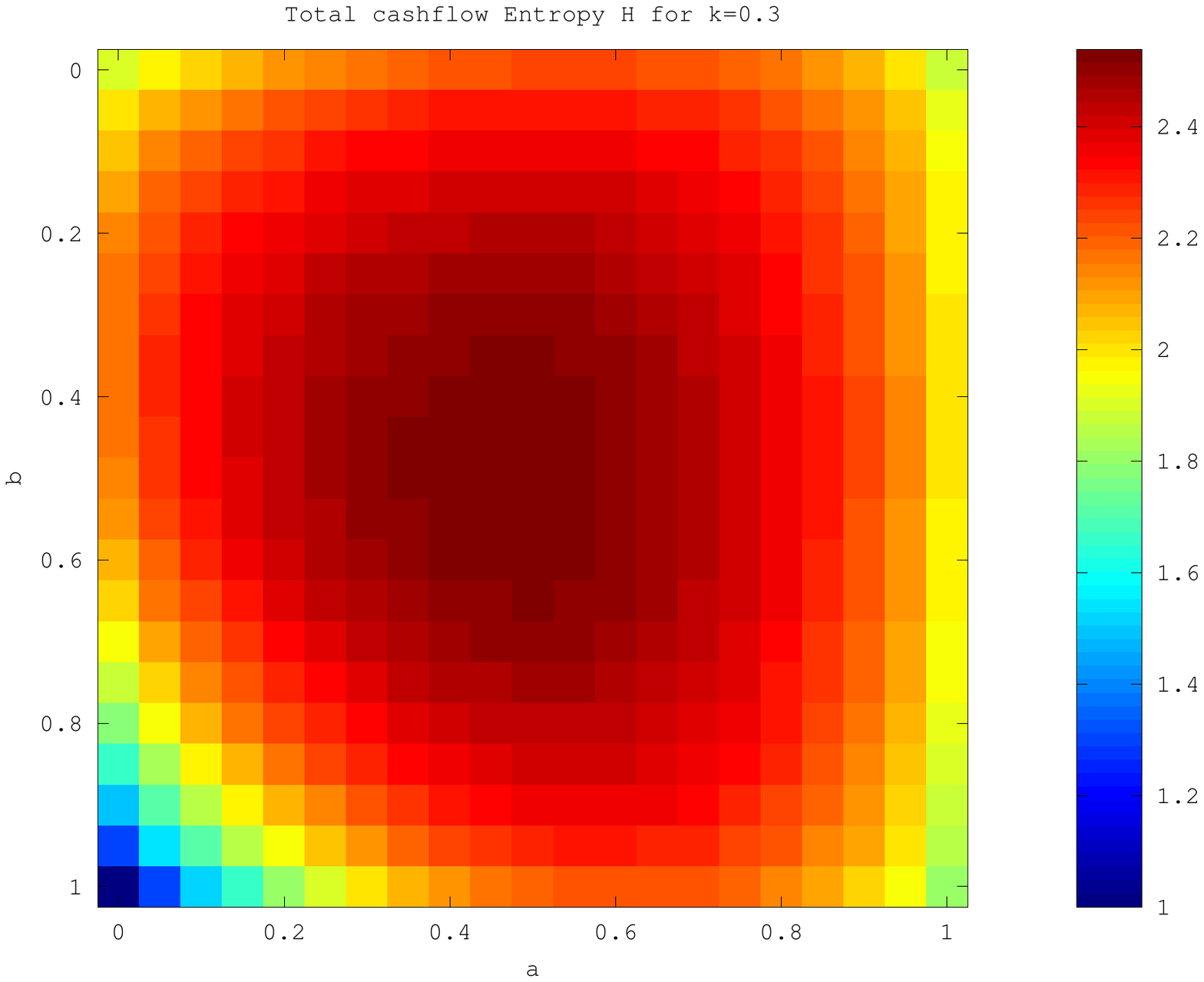}
\end{center}
\caption{
Heat-map of the overall cash-flow entropy $H$ for the 3-agent case with $k=0.3$.}
\label{fig-3D-H-3}
\end{figure}

\begin{figure}
\begin{center}
\mygraphics{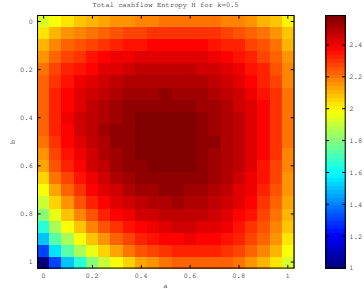}
\end{center}
\caption{
Heat-map of the overall cash-flow entropy $H$ for the 3-agent case with $k=0.5$.}
\label{fig-3D-H-5}
\end{figure}

The overall entropy (which is equal to $H_{\{c\}}$ because we excluded savings here)
for different values of $k$ is presented in the heat plots \ref{fig-3D-H-3} and \ref{fig-3D-H-5}.

To illustrate the identity (\ref{eq-280}) let us assume a specific case, by fixing the parameters for the cash flow matrix
to
\be
a=0.1 \quad b=0.3 \quad k=0.7.
\ee
This corresponds to the cash flow matrix
\be
\left[
\begin{array}{ccc}
0 & 0.1235 & 1.1118 \\
0.3516 & 0 & 0.1507 \\
0.8837 & 0.3787 & 0 
\end{array}
\right]
\ee
The inflow entropies (entropies in each column) are then
\be
H_1=0.8617; \quad H_2 =0.8048; \quad H_3=0.5275
\ee
while the outflow entropies (entropies in each row) are then
\be
H^1=0.4690; \quad H^2=0.8813; \quad    H^3=0.8813.
\ee
Let us define the entropy differentials (inflow minus outflow entropy) $D_j \de H_j-H^j$
\be
D_1=0.3927; \quad D_2=-0.0765; \quad D_3=-0.3538
\ee
We see that agent 1 is an entropy reducer, while agent 3 is an entropy increaser. Agent 2 is almost entropy neutral.
From the cash flow matrix we find the $p_j$ as
\be
p_1=0.4118; \quad    p_2=0.1674; \quad   p_3=0.4208.
\ee
Now it is easy to verify that the identity (\ref{eq-280}) is satisfied.
This means that the weighted entropy differentials for any two agents imply the
weighted entropy differential for the remaining agent.

\section{Conclusion}
\label{sec_conc}
We presented a formalism linking cash flows to an entropy measure. The measure naturally adopts to sub-goupings of the cash flows and
as such produces a number of inequality measures, one of them being the Theil Index.

While the Theil Index is well know, we believe the over all framework we presented is new.
It produces inequality measures between all possible groupings of agents and internal inequality measures for each group.

Another interesting result is a kind of overall entropy balance expressed in equation (\ref{eq-220}) and (\ref{eq-280}).
This essentially says that there must be a balance between entropy increasers (receiving few large cash flow, making many small payments) and decreasers
(receiving many small payments and making a few large payments). Whether this can be linked to an ``economic energy'' is an interesting question which we
have not yet addressed. However, one might speculate that entropy increasers (consumers) tend to pay the profit margins received by entropy decreasers
(companies).

The formalism presented here does not allow for credit, which is obviously an important factor in the real world.
We believe that credit can be incorporated in the formalism and this will be the subject of further research.

Another possible extension is an inclusion of time so that entropy cannot just be measured over one time step, but also
over a number of time steps. One way of doing this is to consider discrete times and to treat the same agent at
different times as 'different'. This should lead to entropies for payment profiles over time, for example for loans
repayed in installments. We hope to be able to explore this more in a seperate paper.

\appendix

\section{Subdivision Identity Example}
Let us assume that we have $2$ possible outcomes labeled by $1$ and $2$ with probabilities $p_1$ and $p_2=1-p_1$.
The information entropy is then
\be
H=-p_1 \log(p_1) - p_2 \log(p_2).
\ee
Now let outcome $2$ be the combination of two possible outcomes $21$ and $22$ with probabilities $p_{21}$ and $p_{22}$ such
that $p_2 = p_{21} + p_{22}$. We then find for the over-all entropy
\begin{eqnarray}
H&=&-p_1 \log(p_1) - p_{21} \log(p_{21}) - p_{22} \log(p_{22}) \nonumber \\
 &=&-p_1 \log(p_1) \nonumber \\
 && \quad - p_{21} \log\left(p_2 \frac{p_{21}}{p_2}\right)
 - p_{22} \log\left(p_2 \frac{p_{22}}{p_2}\right) \nonumber \\
 &=&-p_1 \log(p_1) - p_2 \log(p_2) \nonumber \\
&& \quad + p_2 \left[
- \frac{p_{21}}{p_2} \log\left(\frac{p_{21}}{p_2}\right)
-\frac{p_{22}}{p_2} \log\left(\frac{p_{22}}{p_2}\right)
 \right] \nonumber
\end{eqnarray}
The first two terms are the information entropy for outcome $1$ and the aggregate outcome $2$.
The second term then adds the internal information entropy
resulting from the sub-division of outcome $2$, which is the weighted information entropy of the probabilities conditional on the
knowledge that outcome $2$ is true. 

Above derivation can be generalized in an obvious way to arrive at (\ref{eq-50}).

\end{document}